\documentclass[12pt]{article}

\usepackage{epsfig}

\begin{document}

\def\d{\displaystyle}
\def\be{\begin{equation}}
\def\ee{\end{equation}}
\def\bea{\begin{eqnarray}}
\def\eea{\end{eqnarray}}
\def\gsim{\:\raisebox{-0.75ex}{$\stackrel{\textstyle>}{\sim}$}\:}
\def\lsim{\:\raisebox{-0.75ex}{$\stackrel{\textstyle<}{\sim}$}\:}

\newcommand{\refb}[1]{(\ref{#1})}
\newcommand{\p}{\partial}
\newcommand{\half}{{1\over 2}}
\newcommand{\sectiono}[1]{\section{#1}\setcounter{equation}{0}}
\def\be{\begin{equation}}
\def\ee{\end{equation}}
\def\ba{\begin{array}{l}}
\def\ea{\end{array}}
\def\bea{\begin{eqnarray}}
\def\eea{\end{eqnarray}}
\def\eq#1{(\ref{#1})}
\def\fig#1{Fig \ref{#1}} 
\def\wgnc{\bar{\wedge}}
\def\half{\frac{1}{2}}
\def\nn{\nonumber}
\def\ov{\over}
\def \opr#1{\frac{\partial}{\partial #1}}
\def\vev#1{\langle #1 \rangle}
\newcommand{\bra}[1]{\langle #1|}
\newcommand{\ket}[1]{|#1\rangle}
\newcommand{\inner}[2]{\langle #1|#2\rangle}

\title{String Theory: A Framework for Quantum Gravity
and Various Applications}
\author{Spenta R. Wadia \\[.5cm]
International Center for Theoretical Sciences \\
and \\
Department of Theoretical Physics \\ Tata Institute of Fundamental
Research \\ Homi Bhabha Road, Mumbai 400 005, India \\ 
{\it e-mail: wadia@theory.tifr.res.in}}
\date{}
\maketitle

\begin{center}
\underbar{\large Abstract}
\end{center}
\bigskip

In this semi-technical review we discuss string theory (and all that
goes by that name) as a framework for a quantum theory of gravity.
This is a new paradigm in theoretical physics that goes beyond
relativistic quantum field theory.  We provide concrete evidence for
this proposal. It leads to the resolution of the ultra-violet catastrophe
of Einstein's theory of general relativity and an
explanation of the Bekenstein-Hawking entropy (of a class of black
holes) in terms of Boltzmann's formula for entropy in statistical
mechanics.  We discuss `the holographic principle' and its precise and
consequential formulation in the AdS/CFT correspondence of Maldacena.  One
consequence of this correspondence is the ability to do strong
coupling calculations in $SU(N)$ gauge theories in terms of
semi-classical gravity.  In particular, we indicate a 
connection between dissipative fluid dynamics and the dynamics of
black hole horizons. We end with a discussion of elementary particle
physics and cosmology in the framework of string theory.  We do not
cover all aspects of string theory and its
applications to diverse areas of physics and mathematics, but follow a
few paths in a vast landscape of ideas.

\noindent {\it (This article has been written for TWAS Jubilee
publication).}  
\newpage

\noindent {\Large\bf 1 Introduction:}
\bigskip

In this article we discuss the need for a quantum theory of gravity to
address some of the important questions of physics related to the very
early universe and the physics of black holes.  We will posit the case
for string theory as a framework to address these questions.  The
bonus of string theory is that it has the tenets of a unified theory
of all interactions, electro-magnetism, weak and strong interactions,
and gravitation.
Given this, string theory provides a framework to address some
fundamental issues in cosmology and elementary particle physics.  
Examples are dark matter, supersymmetric particles, dark energy, 
unification for all interactions etc. etc.
\bigskip

Perhaps the most important success of string theory, in recent times,
is in providing a microscopic basis of black hole thermodynamics.  The
discovery of the AdS/CFT correspondence, as a precise realization of
the holographic principle of black hole physics, has shed new light on
the solution of large $N$ gauge theories and other field theories at
strong coupling.  Given the diversity of concepts and techniques, 
string theory has a healthy interface with various branches of
mathematics and statistical mechanics.  More recently there have
appeared connections with fluid mechanics and strongly coupled
condensed matter systems.
\bigskip

We begin with a brief review of the current theories of physics and
their limitations.
\bigskip

\noindent {\Large\bf 2 Quantum Mechanics and General 
\smallskip

Relativity:}
\bigskip

\noindent {\bf Quantum Mechanics}:
\bigskip

Quantum mechanics is the established framework to describe the world
of molecules, 
atoms, nuclei and their constituents.  Its validity has been tested to
very short distances like $10^{-18}$ meters in high energy collision
experiments at CERN and Fermi Lab.  In Quantum mechanics the scale of
quantum effects is set by Planck's constant $\hbar = 1.05 \times 10^{-27}$
erg. sec., and a new mathematical formulation is required.
Position and momentum do not commute $xp - px = i\hbar$, and the 
Heisenberg uncertainty relation $\Delta x \ \Delta p \geq \hbar$
implies a limit to which we can localize the position of a point
particle.  This fuzziness that characterizes quantum mechanics 
resolves e.g. the fundamental problem of the stability of atoms. 
\bigskip

\noindent {\bf Relativistic Quantum Field Theory}:
\bigskip

The application of quantum mechanics to describe relativistic
particles, requires the framework of quantum field theory (QFT),
which is characterized by $\hbar$ and the speed of light $c$.  The
successful theories of elementary particle physics viz. electro-weak
theory and the theory of strong interactions are formulated in this
framework.  The inherent incompatibility between a `continous field' and
the notion of a quantum fluctuation at a space-time point is resolved
in the framework of renormalization theory and the concept of the
renormalization group that was developed by K.G. Wilson \cite{1one}.
\bigskip

\noindent {\bf General Relativity}:
\bigskip

General Relativity was conceived by Einstein to resolve an apparent 
contradiction between special relativity and Newton's theory of
gravitation.  In special relativity, interactions take a finite time
to propagate due to the finiteness of the speed of light.  But in
Newton's theory the gravitational interaction is instantaneous!  The
resolution in general relativity is that space-time is not static and
responds to matter by changing its geometry (the metric of space-time) 
in accordance with
Einstein's equations.  General relativity is a very successful theory
(by success we mean it is experimentally well tested) for distances
$R$ and masses $M$ characterized by $R \gsim \ell_{Pl}$, where
$\ell_{Pl} = 2G_N M/c^2$ ($G_N$ is Newton's constant) is the Planck
length.  This includes a large range of phenomena in relativistic 
astrophysics.  General relativity also provides the framework of the
standard inflationary model of cosmology within which one successfully
interprets the cosmic microwave background (CMB) data.
\bigskip

Now let us indicate the problem one encounters when one tries to
quantize general relativity.  
\bigskip

\noindent {\it Divergent Quantum Theory}:
\bigskip

Just like in Maxwell's theory,
where electro-magnetic waves exist in the absence of sources,
Einstein's equations predict `gravity' waves.  The `graviton' is the
analogue of the `photon'.  The graviton can be considered as a
`particle' in the quantum theory with mass $M = 0$ and spin $S = 2$.
It is a fluctuation of the geometry around flat Minkowski
space-time. 
\bigskip

Given this, one does the obvious like in any quantum field theory.  One
discusses emission and absorption processes of gravitons.  The
non-linearity of Einstein's theory implies that besides the emission
and absorption of gravitons by matter, gravitons can be emitted and
absorbed by gravitons.  These processes are characterized by a
dimensionless `coupling constant': $E/E_{Pl}$, where $E$ is the energy
of the process and $E_{Pl} = (\hbar c^5/G_N)^{1/2} \simeq 10^{19} \
{\rm GeV}$.  As long as we are discussing processes in which
$E/E_{pl} \ll 1$, the effects of quantum fluctuations are
suppressed and negligible.  When $E/E_{pl} \gsim 1$ gravity is
strongly coupled and graviton fluctuations are large.  It is basically
this fact that renders the standard quantum theory of the metric
fluctuations around a given classical space-time badly divergent and
meaningless. Note that in discussing quantum gravity effects, we
introduced the Planck energy which involves all the 3 fundamental
constants of nature: $\hbar, c$ and $G_N$.  
\bigskip

\noindent {\it Big Bang Singularity}:
\bigskip

If we consider the Friedmann-Robertson-Walker (FRW) expanding universe
solution and extrapolate it 
backward in time, the universe would be packed in a smaller and
smaller volume in the far past.  Once again when we reach the Planck
volume $\ell^3_{Pl}$, we would expect a breakdown of the quantized
general relativity, due to large uncontrolled fluctuations.  In the
absence of a theory of this epoch of space-time it would be impossible
to understand in a fundamental way the evolution of the universe after
the `big bang'.  In fact the very notion of a `initial time' may loose meaning.
Perhaps a new framework may provide a new language in terms of
which we may address such questions about the very early universe.
\bigskip

\noindent {\it Black Hole Information Paradox}:
\bigskip

Quantized general relativity also runs into difficulties with quantum
mechanics in the description of phenomena in the vicinity of the
horizon of a black hole.  A black hole is formed when a large mass is
packed in a small volume, characterized by the radius $r_h =
2G_N M/c^2$.  If so then even light cannot escape
from its interior and hence the name black hole.  The surface with
radius $r_h$ is called the horizon of a black hole and it divides the
space-time into 2 distinct regions.  The horizon of the black hole   
is a one way gate: If you check in you cannot get out!  However
Hawking in 1974 realized that in quantum mechanics black holes radiate.  He
calculated the temperature of a black hole of mass $M$: $T =
\hbar c^3/8\pi G_N M$. Using the first law of
thermodynamics and Bekenstein's heuristic proposal that the entropy of
a black hole is proportional to the area of its horizon, he arrived at
one of the most important facts of quantum gravity: The entropy of a
large black hole is given by
\be
S_{bh} = {A_h c^3 \over 4\hbar G_N}
\label{one}
\ee
where $A_h$ is the area of the horizon of the black hole.  In general
the black hole is characterized by its mass, charge and angular momentum
and the Bekenstein-Hawking formula (\ref{one}) is valid for all of them.
Note that it involves all the 3 fundamental constants 
$\hbar, c$ and $G_N$.  Defining $A_{Pl} = \hbar G_N/c^3$ (Planck area)
we can write it suggestively as 
\be
S_{bh} = {A \over 4 A_{Pl}}.
\label{two}
\ee
$A/A_{Pl}$ represents the number of degrees of
freedom of the horizon.  In $3+1$ dim. $A_{Pl} \approx 2.6 \times
10^{-70} \ {\rm m}^2$.  {\it This formula is a bench mark for any theory of
quantum gravity to reproduce.}
\bigskip

The fact that the entropy is proportional to the area of the horizon
and not the volume it encloses, gives a clue that even though the
degrees of freedom of the black hole are apparently behind the
horizon, they seem to leave an imprint (hologram) on the horizon.  
\bigskip

Once the black hole is formed it will emit Hawking radiation.  It is
here that we run into a problem with quantum 
mechanics.  The quantum states that make up the black hole cannot be
reconstructed from the emitted radiation, even in principle, within the
framework of general relativity because the emitted radiation is
thermal.  This is the celebrated `information 
paradox' of black hole physics.  It is clear that the resolution of
this paradox is intimately connected with an understanding of the
Bekenstein-Hawking entropy formula and the degrees of freedom that
constitute the black hole.
\bigskip

\noindent {\bf Why String Theory}?
\bigskip

Now that we have spelt out (in three important instances) why quantizing
general relativity does not produce a theory of quantum gravity, we
would like to posit the view that the correct framework to address the
issues we have raised is String theory.  {\it Unlike the development of
general relativity which had the principle of equivalence as a guide
from the very begining, string theory has no such recogniziable
guiding principle.  But perhaps it may just turn out that the
`holographic principle', which we will discuss later on, is one such,
guiding principle}.
\bigskip

\noindent {\Large\bf 3 String theory primer:\footnote{A concise
modern reference on string theory is the book by E. Kiritsis
\cite{one}.  The classic text books are \cite{two-a} \cite{two-b}.}} 
\bigskip  

\noindent{\large\bf 3.1 Perturbative string theory} 
\medskip

The laws of nature in classical and quantum mechanics are usually
formulated in terms of point particles.  When many particles are
involved, the mechanical laws are described in terms of fields.  A
good example is the Navier-Stokes equation which is Newton's laws
for a fluid.  Even in quantum field theory we essentially deal with
the `particle' concept because in an approximate sense a field at a
point in space creates a particle at that point from the vacuum.  
{\it The main paradigm shift in string theory is that the formulation
of the dynamical laws is not restricted to point particles}.
\bigskip

\begin{center}
\epsfig{file=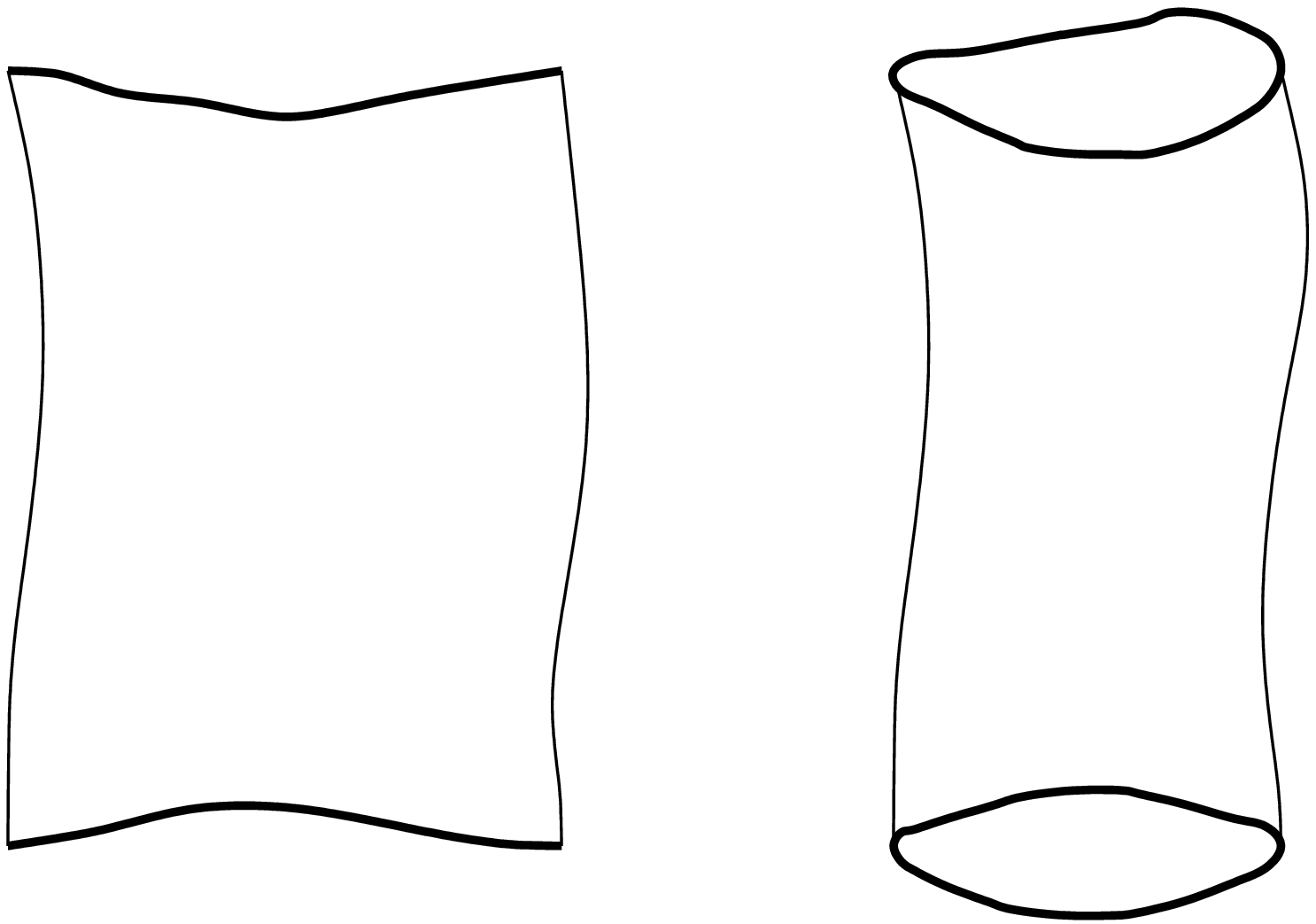,height=5cm}
\end{center}
\centerline{Fig. 1}
\bigskip

Historically the first example beyond point particles is the one
dimensional string.  Strings can be open or closed.  The open string
sweeps out a world sheet with end points moving at the speed of light
(Fig. 1).  The closed string sweeps out a cylindrical surface in 
space-time (Fig. 1).  The dynamics is determined by an action principle which
states that the area swept out between initial and final
configurations is minimum.
\bigskip

String theory comes with an intrinsic length scale $\ell_s$ which is
related to the string tension $T = \ell^{-2}_s$.  One of the great
discoveries of string theory is that a string can carry bosonic as
well as fermionic coordinates: $X^\mu (\sigma,t)$,
$\psi^\mu(\sigma,t)$, $\bar\psi^\mu(\sigma,t)$.  $X^\mu$ is a
space-time coordinate and $\psi^\mu$, $\bar\psi^\mu$ are additional
`anti-commuting' coordinates.  This is a radical extension of our
usual notion of space-time.  The action describing
the free string dynamics is  
\be
S = {1 \over \ell^2_s} \int d\sigma \ dt\left(\partial_\alpha X^\mu \
\partial^\alpha X_\mu - i \bar\psi^\mu \rho^\alpha \partial_\alpha \
\psi_\mu\right) 
\label{three}
\ee
The index `$\mu$' is a space-time index, `$\alpha$' is a 2-dim. world
sheet index and $\rho^\alpha$ are 2-dim. Dirac matrices. Besides the
standard conformal symmetry this action is invariant under a
supersymmetry transformation\footnote{Supersymmetry was discovered 
first in string theory by Ramond, Neveu, Schwarz, and Gervais and
Sakita.  This later 
inspired the construction of supersymmetric field theories in $3+1$
dim. \cite{two}.} which transforms fermions into bosons and vice versa:
\be
\delta X^\mu = \bar\epsilon \psi^\mu, \ \delta \psi^\mu =
-i\rho^\alpha \partial_\alpha X^\mu \epsilon.
\label{four}
\ee
The symmetry transformation parameter $\epsilon$ is a constant anti-commuting
spinor.  The string described by (\ref{three}) is called a superstring. 
\bigskip

\noindent {\bf String Spectrum}:
\bigskip

The various vibrational modes of the superstring correspond to an
infinite tower of particle states of spin $J$ and mass $M$, satisfying
a linear relation $M^2 = \displaystyle{1 \over \ell^2_s} J +$ const.
It turns out that the spectrum of the superstring can be organized
into space-time supersymmetry multiplets.  {\it The most important aspect of
the spectrum of the string is that in flat 10 space-time dimensions (and
none other), 
its spectrum has a graviton and a gluon.  The graviton $(J =
2,M=0)$ is the massless particle of the closed string, while the gluon
$(J=1,M=0)$ is the massless particle of the open string.  Their
space-time supersymmetric partners are the gravitino $(J = 3/2,M=0)$
and the gluino $(J= 1/2,M=0)$}.  
\bigskip

\noindent {\bf String Interactions}:
\bigskip

String interactions involve the spliting and joining of strings,
characterized by a coupling constant denoted by $g_s$, the string
coupling (Fig. 2a).  

\begin{center}
\epsfig{file=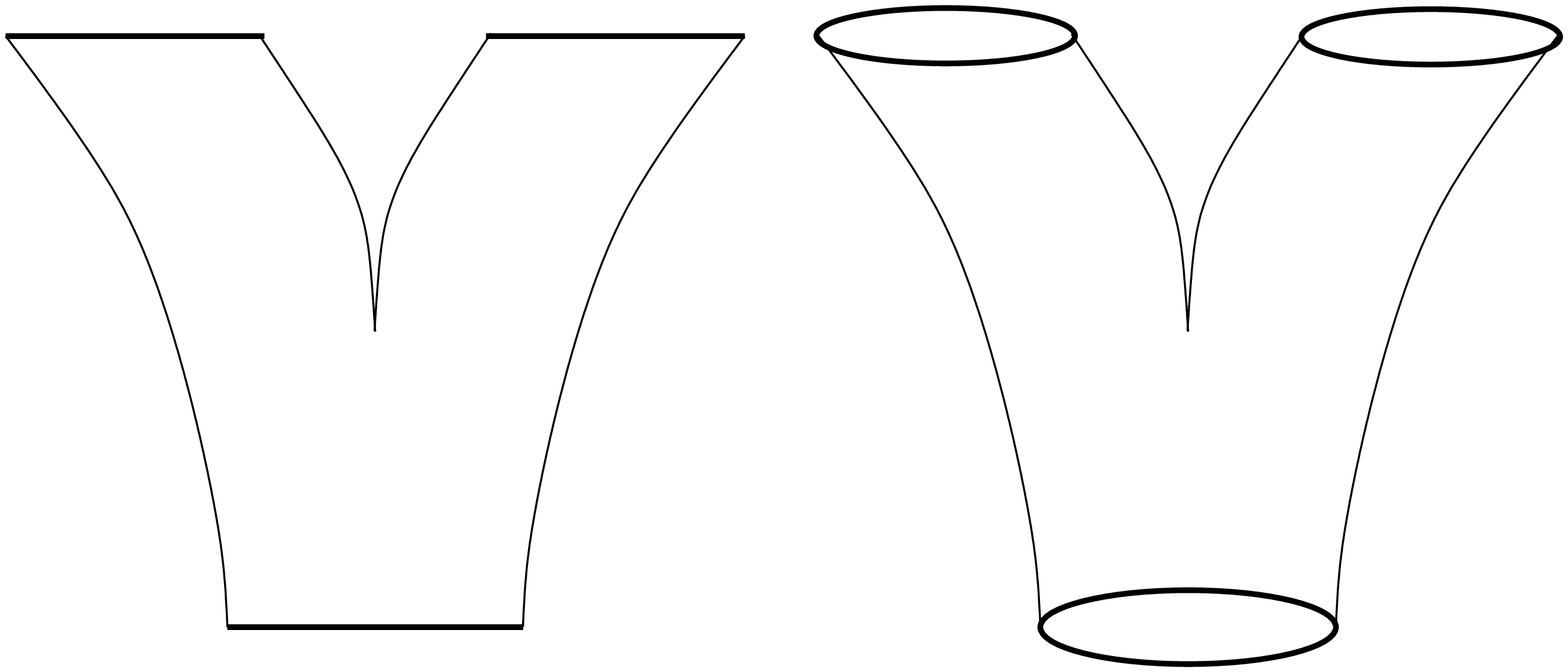,height=4cm}
\centerline{Fig. 2a}
\end{center}

\noindent These interactions generate 2-dim. surfaces with
non-trivial topology.  For example the spliting and rejoining of a
closed string creates a handle on the world sheet.  A similar process
for the open string creates a hole in the world sheet.  There are also
diagrams describing the interactions of closed and open strings (Fig. 2b).
These interactions can be consistently described only in a
10-dim. space-time. 

\begin{center}
\epsfig{file=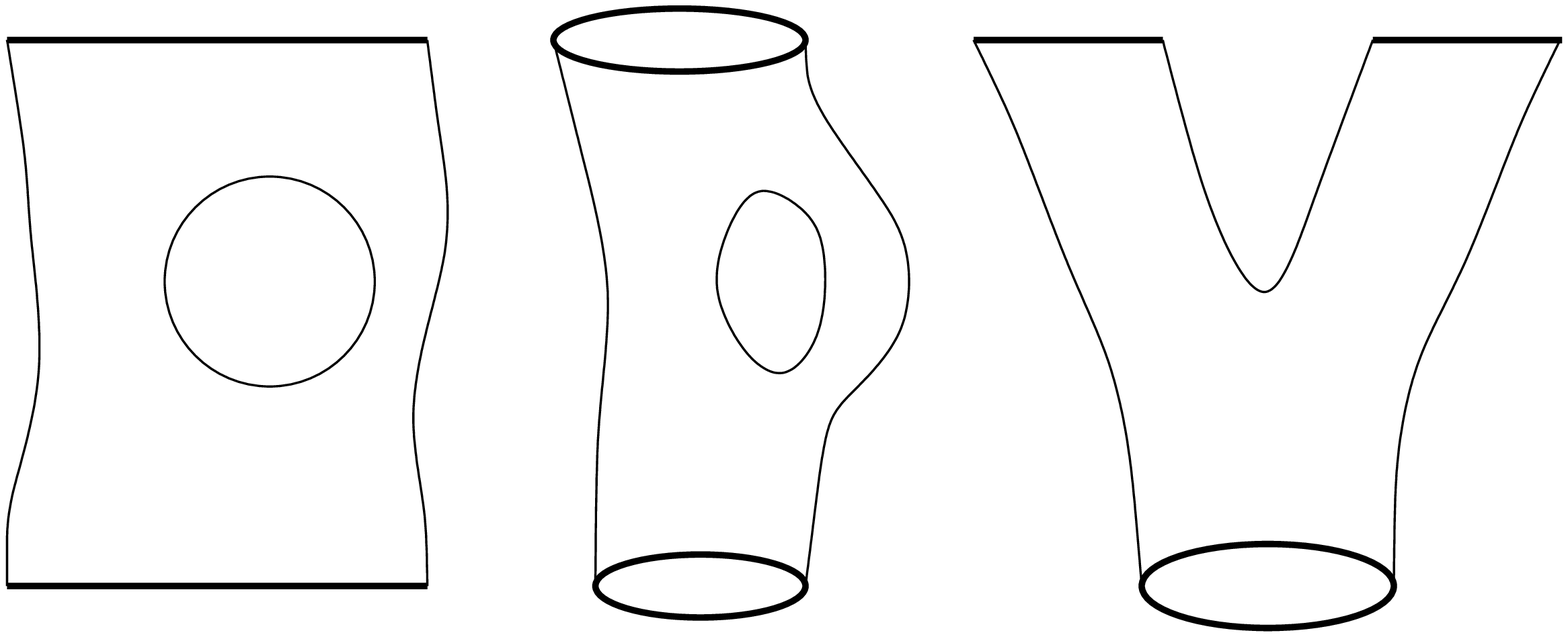,height=4cm}
\end{center}
\centerline{Fig. 2b}
\bigskip

Now the rules of perturbative string theory which we have briefly 
stated enable us to calculate scattering of the particle states of
the free string theory.  In particular graviton scattering amplitudes
in string theory turn out (to leading order in the $E/M_{Pl} c^2$)
to be identical to those calculated from general relativity (actually
supergravity) with the identification of the 10-dim.  Newton's
constant \cite{yoneya} \cite{sherk} 
\be
G_N^{(10)} = 8\pi^6 \ell^8_s g^2_s.
\label{five}
\ee

The deep result here is that unlike general relativity string theory has no
short distance divergences.  This is because in general relativity
point-like gravitons with large energy come arbitrarily close together
in accordance with the uncertainty principle $\Delta x \geq
\hbar/\Delta p$.  In string theory due to the
existense of a length scale $\ell_s$, high energy gravitons do not
come arbitrarily close together (see Fig. 2a), but the energy gets
distributed in the 
higher modes of the string, so that instead of probing short distances
the size of the string grows.  This point can be summarized in a
plausible generalization of the Heisenberg uncertainty principle 
\be
\Delta x \geq {\hbar \over \Delta p} + \ell^2_s {\Delta p \over \hbar}
\label{six}
\ee 
{\it Hence string theory by its construction, includes an additional
length scale $\ell_s$, and gives rise to a ultra-violet finite theory.
This result is radically different from standard quantum field
theory.  In this way string theory passes its first test towards a
theory of quantum gravity: it is perturbatively finite and
calculable}. 
\bigskip

\noindent {\bf 5 Types of String Theories in 10-dim.}:
\bigskip

Various perturbatively consistent superstrings have been constructed
in 10 space-time dimensions.  There are 5 such theories:
\begin{enumerate}
\item[{(i)}] Type I strings are open and can have non-abelian charges
at their ends.  These open strings can also evolve into closed Type I
strings.  (Open Type I strings are analogous to QCD strings
with quarks at their end points. Closed Type I strings are 
analogous to closed QCD strings which describe gluons.)
\item[{(ii)}] Type II A and Type II B in which the strings are closed
and oriented.
\item[{(iii)}] The 2 Heterotic string theories corresponding to the
gauge groups $SO(32)$ and $E_8 \times E_8$.
\end{enumerate}
\bigskip

\noindent{\large\bf 3.2 Non-perturbative string theory} 
\bigskip

\noindent (i) \underbar{Duality and $M$-theory}:
\bigskip

Going beyond a perturbative description of string theory is rendered
difficult by the fact that unlike in standard quantum field 
theory, string theory is defined by a set of rules that enable a perturbative
definition.  However progress was made towards understanding
non-perturbative effects in string theory by the discovery of duality
symmetries. 
\bigskip

The duality symmetries of string theory are of 2-kinds.  One called
$S$-duality, has a correspondence in standard quantum field theory.
It says that 2 theories with coupling constants $e$ and $g$ are
actually the same theory when $eg = c$, a constant.  This implies that
the theory with coupling constant $e$, can be studied at strong
coupling $(e \gg 1)$, by the equivalent version with coupling constant
$g = c/e \ll 1$.  
Examples are the Sine-Gordon and Thirring model, $Z_2$ gauge theory
and the Ising model in $2+1$ dim., and Maxwell's theory in $3+1$ dim.
In statistical physics this duality is called the Kramers-Wannier
duality.  In the case of Maxwell's theory it is called
`Electric-Magnetic' duality, where the electric and magnetic charges
are related by $eg = \hbar/2$.
\bigskip

The other duality symmetry called $T$-duality has no analogue in
quantum field theory and is stringy in origin \cite{three}.  An
example is a string moving in a space-time, one of whose dimensions is
a circle of radius $R$. It is easy to see that the energy levels are
given by (assuming motion is only along the circle) 
\be E_n =
{c\hbar|n| \over R} + {c\hbar |m| R \over \ell^2_s}
\label{seven2}
\ee
where $\hbar n/R$ 
is the momentum of the state characterized by the integer $n$ and
$c\hbar |m| R/\ell_s^2$ is the energy due to the winding modes of the
string.  The energy formula is symmetric under the interchange of
momentum and winding modes $(|n| \rightarrow |m|)$ and $R
\rightarrow \ell^2_s/R$.  For $R \gg \ell_s$ the momentum mode description is
appropriate while for $R \ll \ell_s$ the winding mode
description is more appropriate.  At $R = \ell_s$ both
descriptions are valid.  This simple example also generalizes to
higher dimensional objects called `branes'. 
\bigskip

Duality has a profound consequence for string theory.  $T$-duality led
to the inference of $D$-branes, which are solitonic domain walls on
which open strings end \cite{four} (that explains the $D$, because the
open string has 
a Dirichelet boundary condition on the brane).  {\it A combination of $S$
and $T$ dualities led to the realization that the 5 perturbatively
defined string theories are different phases of a meta-theory called
$M$-theory which lives in 11-dim. The radius of the 11th
dimension $R_{11} = \ell_s g_s$, so that as $g_s \rightarrow \infty$,
$R_{11} \rightarrow \infty$.  One of the great challenges of
string theory is to discover `new principles' that would lead to a
construction of $M$-theory.}
\bigskip

\noindent (ii) \underbar{$D$-branes}:
\bigskip

We now give a brief introduction to $D$-branes as they, among other
things, play a crucial
role in resolving some of the conundrums of black hole physics.  As we
have mentioned a $D$-$p$ brane (in the simplest geometrical
configuration) is a domain wall of dimension $p$, where $0
\leq p \leq 9$.  It is  
characterized by a charge and it couples to a $(p+1)$ form
abelian gauge field $A^{(p+1)}$\footnote{e.g. a D0 brane couples to a
1-form gauge field $A^{(1)}_\mu$, a D1 brane couples to a 2-form
gauge field $A^{(2)}_{\mu\nu}$ etc.}.  The $D$-$p$ brane has a brane
tension $T_p$ which is its mass per unit volume.  The crucial point is
that $T_p \propto 1/g_s$.  This dependence on the coupling
constant (instead of $g^{-2}_s$) is peculiar to string theory.  It has
a very important consequence.  A quick estimate of the gravitational
field of a $D$-$p$ brane gives, $G_N^{(10)} T_p \sim g^2_s/g_s \sim 
g_s$.  Hence as $g_s \rightarrow 0$, the gravitational field goes to
zero! If we stack $N$ $D$-$p$ branes on top of each other then the
gravitational field of the stack $\sim Ng_s$.  A useful limit to study
is to hold $g_s N = \lambda$ fixed, as $g_s \rightarrow 0$ and $N
\rightarrow \infty$.  In this limit when $\lambda \gg 1$ the stack of
branes can source a solution of supergravity. On the other hand when
$\lambda \ll 1$ there is also a 
description of the stack of $D$-branes in terms of open strings.  
A stack of $D$-branes interacts by the exchange of open strings very
much like quarks interact by the exchange of gluons.  Fig. 3a
illustrates the self-interaction of a D2-brane by the emission and
absorption of an open string and Fig. 3b illustrates
the interaction of 2 D2-branes by the exchange of an open string.

\begin{center}
\epsfig{file=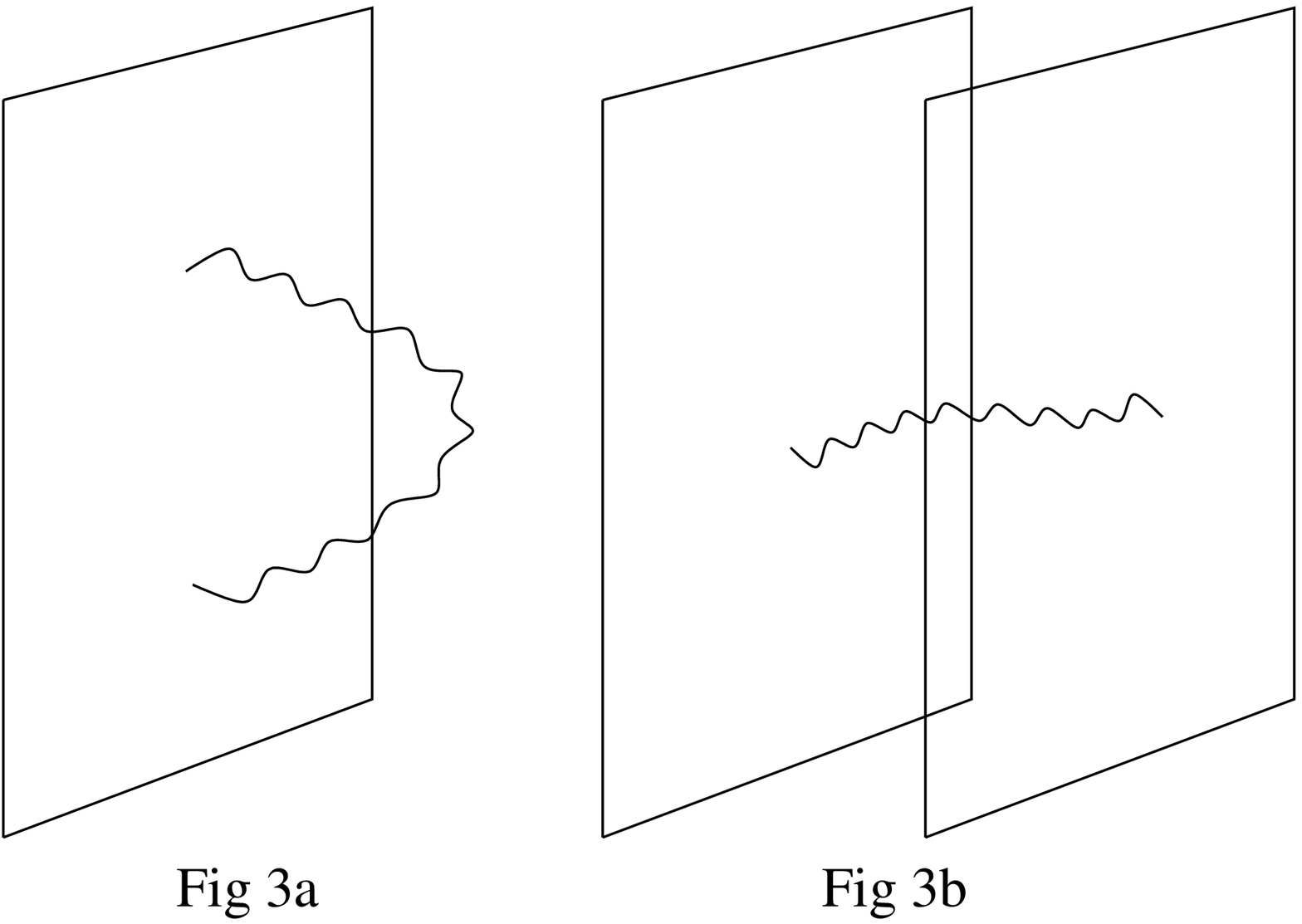,height=5cm}
\end{center}
\centerline{Fig. 3}

\noindent In the infra-red limit only the lowest mode of the open
string contributes and hence the stack of $N$ $D$-branes can be
equivalently described as a familiar $SU(N)$ non-abelian gauge theory
in $p+1$ dim.
\bigskip

A precise formulation of the dual description of a large stack of
$D$-branes in terms of gauge fields (or in general a field theory)
and gravity, gives a realization of the holographic principle of black
hole physics.  More on this later.
\bigskip

\noindent{\Large\bf 4 Black hole micro-states:\footnote{For a review
see \cite{one,five,six}}}   
\bigskip

In order to discuss the question of the micro-states of a black hole
it will be best to `construct' a black hole whose states one can
count.  This is exactly what Strominger and Vafa \cite{four-b} did.
They considered 
a system of $Q_1$ D1 branes and $Q_5$ D5 branes and placed them on 
5 circles.  The radius $R_5 \gg \ell_s$ and $R_i \sim
\ell_s$, $i = 1,2,3,4$.  This system of branes interacts by the
exchange of open strings and interacts with gravitons around the flat
10-dim. space-time.  If we are interested in the long wave length
dynamics (compared to $\ell_s$) of the D1-D5 system, one can ignore the
gravitons and also the massive modes of the open strings with masses
$\propto 1/\ell_s$.  What one is left with is a
non-abelian gauge theory of the lowest lying modes of the open strings
exchanged between the D1 and D5 branes.

\begin{center}
\epsfig{file=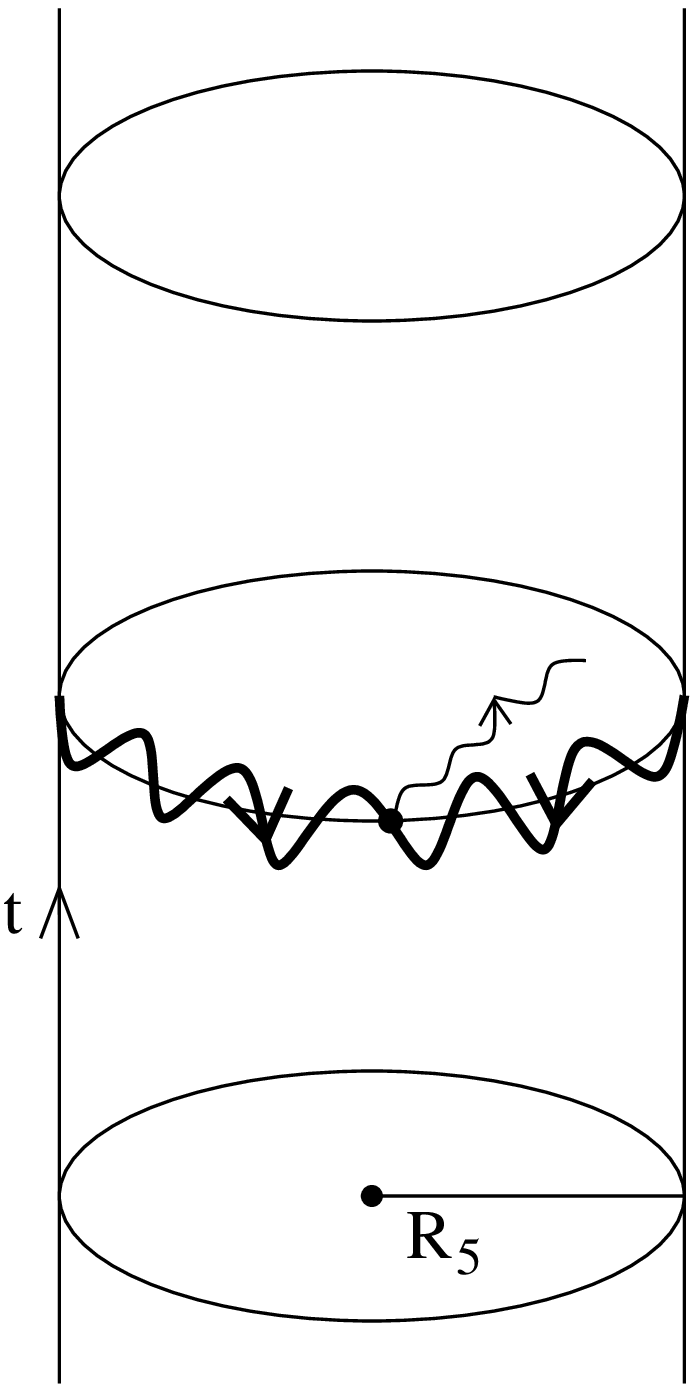,height=6cm}
\end{center}
\centerline{Fig. 4}
\bigskip

These open strings are described by a super-conformal field theory
(SCFT) on a 2-dim. cylinder of radius $R_5$.  Time flows along the
length of the cylinder (Fig. 4).  The central charge of the SCFT
is $C = 6Q_1Q_5$ which basically is the number of open strings (both
bosonic and fermionic) that are exchanged between the D1 and D5
branes.  The SCFT can have excitations moving around the circle.
These can be either left moving or right moving waves.
\bigskip

Now consider a state in the SCFT consisting of these left/right moving
waves with energy $E$ and momentum $P$ (Note that momentum on the
circle is quantized and $P = \hbar n/R_5$, where $n$ is an integer).
There is a 
well known formula due to Cardy which enables us to calculate the
degeneracy of states $\Omega (E,P)$ for fixed values of $E$ and $P$,
for large values of the central charge,
\be
\ln \Omega(E,P) = 2\pi\sqrt{Q_1Q_5} \left(\sqrt{(E + cP)R_5
\over 2\hbar c} + \sqrt{(E - cP)R_5 \over 2\hbar c}\right) +
o\left(\sqrt{Q_1Q_5}\right)
\label{seven}
\ee
(the `$c$' in the above formula is the speed of light).
\bigskip

From here we can calculate the entropy and temperature using
Boltzmann's formulas $S = \ln \Omega$ and $T^{-1} =
\displaystyle{\partial S \over \partial E}$.  In particular in the
case $E = cP$ we get the famous formula
\be
S = 2\pi \sqrt{Q_1Q_5 n}
\label{eight}
\ee
where $E = cP = n/R_5$.  
Now one can find the super-gravity solution corresponding to this
system of branes.  It is a black hole in 4+1 dim. which is
asymptotically, a flat 4+1 dim. space-time.  The brane sources living
in the higher dimensions appear as point sources in 4+1 dim.  The
entropy and temperature of the black hole are calculated using the
Bekenstein-Hawking formula,  
\be
S = {A_h c^3 \over 4G_N \hbar} 
\label{nine}
\ee
with 
\be
A_h = 2\pi^2 r^3_h, \ \ \ \ \ r^2_h = \left(gQ_1 gQ_5 \d{g^2 n \over
R^2_5}\right)^{1/3}
\label{ten}
\ee 
and
\be
G^{-1}_N = \d{2\pi R_5 (2\pi)^4 \over 8\pi^6 g^2_s}
\label{eleven}
\ee is the 5-dim. Newton constant.  We see an exact matching of the
entropy calculated in string theory and general relativity.
\bigskip

The impressive agreement of (\ref{nine}) and the microscopic counting
of black hole entropy continues to hold when we include higher order
corrections (in $\ell_s$) to general relativity.  In this case
(\ref{nine}) was generalized by Wald to be consistent with the first
law of thermodynamics.  Here too the string theory answer, exactly
agrees with Wald's formula.  This agreement is a `precision test' of
string theory as microscopic theory underlying general relativity
\cite{seven}. 
\bigskip

\noindent {\bf Hawking radiation}:
\bigskip

The next important question is whether the microscopic theory can
describe Hawking radiation and predict the decay rate of a black
hole.  For the D1-D5 system the answer to this question is in the
affirmative.  Modeling the emission of Hawking radiation is
similar to the way we describe the emission of a photon from an atom
which is a bound state of electrons and the nucleus.
\bigskip

In the case of the D1-D5 system, the left and right moving waves can
collide to form a closed string mode (Fig.4).  One can calculate an
`$S$-matrix' from an initial state of open strings (left/right moving
waves in the SCFT) to a closed string mode.  The absorption and decay
probabilities are calculated using standard formulas of quantum
statistical mechanics.
\bigskip

In a nutshell consider a microcanonical ensemble ${\cal S}$ specified
by the energy and a set of charges.  Consider the process of
absorption of some particles by the brane system which changes the
energy and charges corresponding to another microcanonical ensemble
${\cal S}'$.  The ensembles ${\cal S}$ and ${\cal S}'$ have total
number of states $\Omega$ and $\Omega'$ respectively.  The absorption
probability from a state $|i\rangle \in {\cal S}$ to a state
$|f\rangle \in {\cal S}'$ is given by
\be
P_{abs} (i \rightarrow f) = {1 \over \Omega} \sum_{i,f} |\langle f |S|
i\rangle|^2
\label{twelve}
\ee
The sum in (\ref{twelve}) is over all final states and there is an
average over all initial states.  Similarly the decay probability is
given by 
\be
P_{\rm decay} (i \rightarrow f) = {1 \over \Omega'} \sum_{i,f}
|\langle f |S| i\rangle|^2
\label{thirteen}
\ee
These formulas then enable us to calculate the decay rate of Hawking
radiation which is given by the formula
\[
\Gamma_H = {\sigma_{abs} (\omega) \over e^{\omega/T_H} - 1} {d^4 k
\over (2\pi)^4}
\]
where $\omega = |k|$, $T_H$ is the Hawking temperature and
$\sigma_{abs} (\omega)$ is the absorption cross section of a given
species of particles at frequency $\omega$.  For spherically symmetric
waves $\sigma_{abs} (\omega \rightarrow 0) = A_h$ the area of the
horizon of the black hole.
\smallskip

The formulas calculated from $D$-branes and gravity match exactly with
those calculated from semi-classical gravity.  The 
important point is that $S_{fi}$ is a standard $S$-matrix and
thermodynamics emerges as an averaging process over microstates of a
micro-canonical ensemble.
\bigskip

In this way the microstate model does indeed explain how Hawking
radiation emerges from a unitary theory.  However the `information
paradox' remains, because its resolution would need us to explain why 
there is loss of unitarity in semi-classical general relativity.
More precisely, even though semi-classical general relativity leads to
the correct formulas for black hole entropy and Hawking radiation
rates, consistent with the laws of thermodynamics, it seems to lack
the ingredients to obtain an in-principle unitary answer without being
embedded in the larger framework of string theory.  
\bigskip

\noindent{\Large\bf 5 Lessons from the D1-D5 system: Holography 

and the AdS/CFT correspondence}{\Large:\footnote{For a review see
\cite{one,three}}}  
\bigskip

We have briefly explained in the preceding section that
thermodynamical properties of black holes and Hawking radiation are
exactly derivable from the dynamics of a stack of D1-D5 branes which
then constitute the microstates of the black hole.  On closer
examination it turns out that these results (on the black hole side)
are determined entirely by the near horizon region of the black hole.
{\it This suggests a general fact that the infrared dynamics of the brane
system is equivalent to gravitational physics in the near horizon
region of the black hole} \cite{eight}.
\bigskip

These suggestive facts about the D1-D5 system, combined with the
`holographic principle' led Maldacena to precisely formulate the
AdS/CFT conjecture.  This is a duality of a non-gravitational theory
with a theory of superstrings (which has supergravity as its low
energy limit).  It gives a precise formulation of the `holographic
principle' (we shall explain this a bit later). It is
simplest to explain this set of ideas in the context of a stack of $N$
D3 branes to which we now turn.
\bigskip

\noindent{\Large\bf 6 D3 branes and the AdS/CFT} 

{\Large\bf correspondence} \cite{one,six}:  
\bigskip

A D3 brane is a 3+1 dim. object.  A stack of $N$ D3 branes interacts
by the exchange of open strings (Fig. 5).  In the long wavelength limit
$(\ell_s \rightarrow 0)$, 

\begin{center}
\epsfig{file=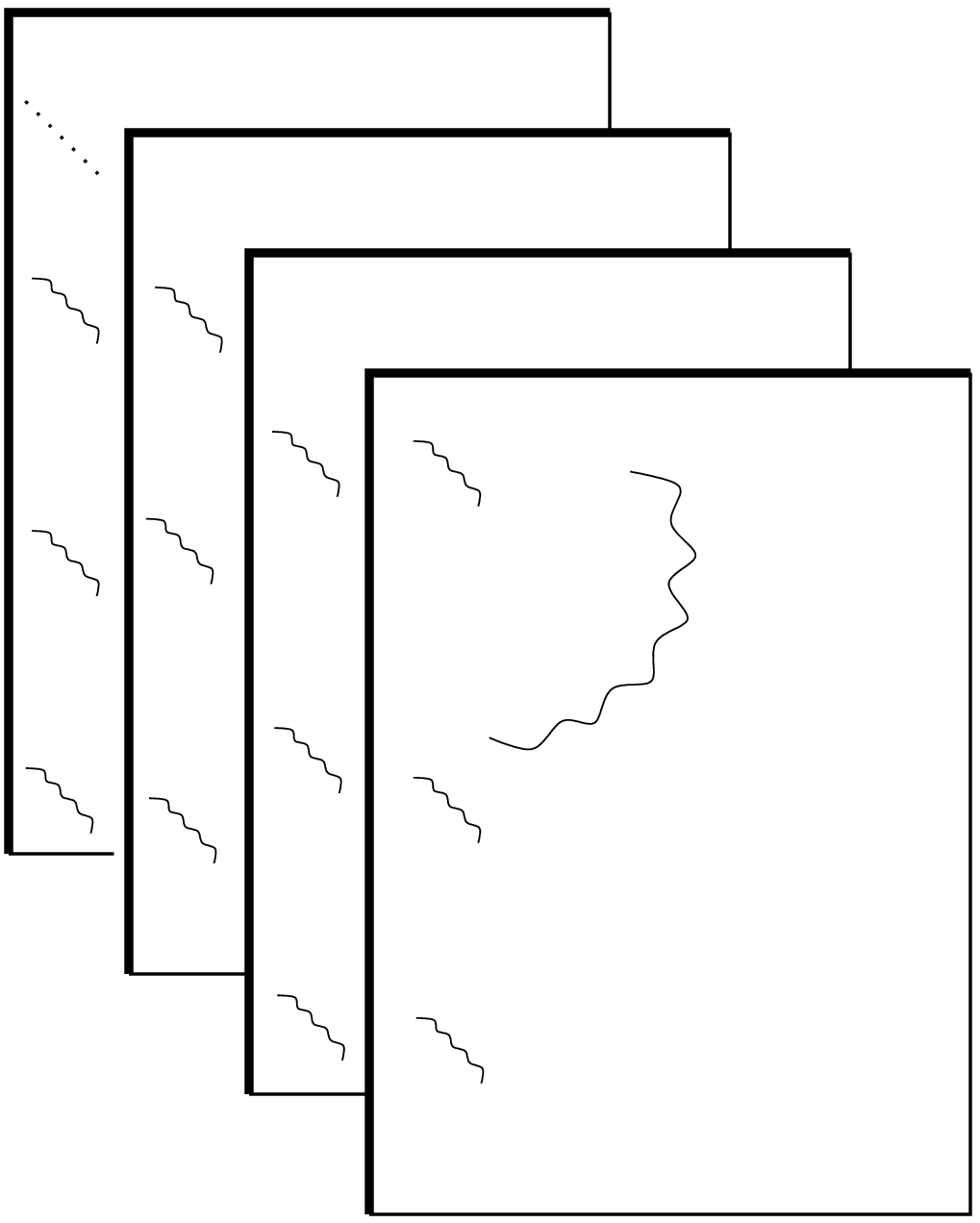,height=6cm}
\end{center}
\centerline{Fig. 5: A stack of $N$ D3 branes interacting via open strings}
\bigskip

\noindent only the massless modes of the open string
are relevant.  These 
correspond to 4 gauge fields $A_\mu$, 6 scalar
fields $\phi^I \ (I = 1,\cdots,6)$ (corresponding to the fact that the brane
extends in 6 transverse dimension) and their supersymmetric partners.
These massless degrees of freedom 
are described by ${\cal N} = 4, \ SU(N)$ Yang-Mills
theory in 3+1 dim.  This is a maximally supersymmetric, conformally
invariant SCFT in 3+1 dim.  The coupling constant of this gauge theory
$g_{YM}$, is simply related to the string coupling $g_s = g^2_{YM}$.
The 'tHooft coupling is $\lambda = g_s N$ and the theory admits a
systematic expansion in $1/N$, for fixed $\lambda$.  Further as
$\ell_s \rightarrow 0$ the coupling of the D3 branes to gravitons
also vanishes, and hence we are left with the ${\cal N}=4$ SYM theory and free
gravitons. 
\bigskip

On the other hand when $\lambda \gg 1$, $N$ D3 branes (for large $N$)
source a 
supergravity solution in 10-dim.  The supergravity fields include the
metric, two scalars, two 2-form potentials, and a 4-form potential whose
field strength $F_5$ is self-dual and proportional to the volume form
of $S^5$.  The fact that there are $N$ D3 branes is expressed as
$\d{\int_{S^5}} F_5 = N$.  There are also fermionic fields required by
supersymmetry.  It is instructive to write down the supergravity metric:
\bea
ds^2 &=& H^{-1/2} (-dt^2 + d\vec x \cdot d\vec x) + H^{1/2} (dr^2 + r^2
d\Omega^2_5) \nonumber \\[2mm]
&& \\
H &=& \left(1 + {R^4 \over r^4}\right), \ \left({R \over
  \ell_s}\right)^4 = 4\pi g_s N \nonumber
\label{fourteen}
\eea
Since $|g_{00}| = H^{-1/2}$ the energy depends on the 5th coordinate
$r$.  In fact the energy at $r$ is related to the energy at $r =
\infty$ (where $g_{00} = 1$) by $E_\infty = \sqrt{|g_{00}|} E_r$.  As
$r \rightarrow 0$ (the near horizon limit), $E_\infty = {r \over R}
E_r$ and this says that $E_\infty$ is red-shifted as $r \rightarrow
0$.  We can allow for an arbitrary excitation energy in string units
(i.e. arbitary $E_r\ell_s$) as $r \rightarrow 0$ and $\ell_s
\rightarrow 0$, by holding a mass scale `$U$' fixed:  
\be
{E_\infty \over \ell_s E_r} \cong {r \over \ell^2_s} = U
\label{fifteen}
\ee
Note that in this limit the bulk gravitons also decouple from the
near horizon region.  This is the famous
near horizon limit of Maldacena and in this limit the metric
(\ref{fourteen}) becomes
\be
ds^2 = \ell^2_s \left[{U^2 \over \sqrt{4\pi\lambda}} \left(-dt^2 +
d\vec x \cdot d\vec x\right) + 4\sqrt{4\pi\lambda} {dU^2 \over U^2}
+ \sqrt{4\pi\lambda} d\Omega^2_5\right]
\label{sixteen}
\ee
This is the metric of ${\rm AdS}_5 \times {\rm S}^5$.  ${\rm AdS}_5$
is the anti-de Sitter space in 5 dim.  This space has a boundary at $U
\rightarrow \infty$, which is conformally equivalent to 3+1
dim. Minkowski space-time.

\newpage

\noindent{\large\bf 6.1 The AdS/CFT conjecture} \cite{eight,six,gubser,nine}
\bigskip

{\it The conjecture of Maldacena is that ${\cal N} = 4$, $SU(N)$ super
Yang-Mills theory 
in 3+1 dim. $\!\!$ is dual to type IIB string theory with ${\rm AdS}_5
\times S^5$ boundary conditions}.  
\bigskip

The gauge/gravity parameters are
related as $g^2_{YM} = g_s$ and $R/\ell_s = (4\pi g^2_{YM} 
N)^{1/4}$.  It is natural to consider the $SU(N)$ gauge theory living
on the boundary of ${\rm AdS}_5$.  The gauge theory is conformally
invariant and its global exact symmetry $SO(2,4) \times SO(6)$, is
also an isometry of ${\rm AdS}_5 \times {\rm S}^5$.  In order to have
a common definition of time in the gauge theory and AdS, the gauge
theory is defined on $S^3 \times R^1$ which is the conformal boundary
of AdS$_5$ and which is conformally equivalent to $R^3 \times R^1$.
Since $S^3$ is compact the gauge theory has no infrared divergences
and hence it is well defined.
\bigskip

The AdS/CFT conjecture is difficult to test because at $\lambda \ll 1$
the gauge theory is perturbatively calculable but the string theory is
defined in ${\rm AdS}_5 \times S^5$ with $R \ll \ell_s$.  On the other
hand for $\lambda \gg 1$, the gauge theory is strongly coupled and
hard to calculate.  In this regime $R \gg \ell_s$ the string
theory can be approximated by supergravity in a derivative expansion
in $\ell_s/R$.  The region $\lambda \sim 1$ is most
intractable as we can study neither the gauge theory nor the string
theory in a reliable way.
\bigskip

\noindent {\bf Interpretation of the radial direction of AdS}: 
\bigskip

Before we discuss the duality further we would like to explain the
significance of the extra dimension `$r$'.  Let us recast the ${\rm
AdS}_5$ metric by a redefinition: $u = \d{R^2 \over r}$
\be
ds^2 = {R^2 \over u^2} \left(-dt^2 + d\vec x \cdot d\vec x +
du^2\right) + R^2 d\Omega^2_5
\label{seventeen}
\ee
The boundary in these coordinates is situated at $u = 0$.  Now this
metric has a scaling symmetry. For $\alpha > 0$, $u \rightarrow \alpha
u$, $t \rightarrow 
\alpha t$ and $\vec x \rightarrow \alpha \vec x$, leaves the metric
invariant.  From this it is
clear that the additional dimension `$u$' represents a length scale in
the boundary space-time: $u \rightarrow 0$ corresponds to a
localization or short distances in the boundary coordinates $(\vec
x,t)$, while $u \rightarrow \infty$ represents long distances on the
boundary. 
\bigskip

Another indicator that the 5th dimension represents a scale in the
gauge theory on the boundary is provided by the fact that the
$r/\ell^2_s = U$, is the energy of an open string 
connecting the stack of $N$ D3 branes and a single D3 brane placed at
a distance `$r$' from it.  In the gauge theory this represents an 
expectation value of the scalar field $\phi^I$ which corresponds to
the symmetry breaking $U(N+1) \rightarrow U(N) \times U(1)$. 
\bigskip

\noindent {\bf Holography}:
\bigskip

We now indicate why the AdS/CFT correspondence gives a holographic
description of physics in AdS.  In order to see this (following
Susskind and Witten) we recast the
metric (\ref{sixteen}) to another form by a coordinate transformation 
\be
ds^2 = R^2 \left[-\left({1 + r^2 \over 1 - r^2}\right)^2 dt^2 + {4
\over (1-r^2)^2} (dr^2 + r^2d\Omega^2)\right] 
\label{eighteen}
\ee
so that the boundary of AdS is at $r=1$.  If we calculate the entropy
of AdS using the Bekenstein-Hawking formula we will get infinity.
Hence we stay near the boundary and set $r = 1 - \delta$, $\delta$ is
a small and $\delta > 0$.  The entropy can now be computed.  An
elementary calculation gives
\be
S = {{\rm Area} \over 4G_N} \cong {R^8 \delta^{-3} \over 4G_N} \sim
{R^8 \delta^{-3} \over g^2_s \ell^8_s} \sim N^2 \delta^{-3}
\label{ninteen}
\ee
When $r = 1 - \delta$, $\delta$ is an ultra-violet cut off of the
boundary theory because as $\delta \rightarrow 0$, the induced metric
on the boundary is 
\be
ds^2 = {R^2 \over \delta^2} \left(-dt^2 + d\vec x \cdot d\vec x\right) 
\label{twenty}
\ee
Now we can easily estimate the degrees of freedom of the $SU(N)$ gauge
theory on $S^3 \times R^1$.  Since $S^3$ is compact, the number of
cells into which we can divide it is $\delta^{-3}$ and hence $S \sim
N^2 \delta^{-3}$.  Hence the estimate of the number of degrees of
freedom in the gauge theory and AdS matches.  This is in accordance
with the holographic principle which states that in a quantum theory
of gravity the degrees of freedom and their interactions can be
described in terms of a non-gravitational theory living on the
boundary of the space-time.
\bigskip

This principle due to 'tHooft and elaborated by Susskind was motivated
by Bekenstein's entropy bound.  Bekenstein argued that the maximum
entropy of a system in a region of space-time is $S = \d{A \over
4G_N}$, where $A$ is the area of the boundary bounding the region.  To
show this assume that this is not true and that it is possible to have
a state in the region with $\bar S > \d{A \over 4G_N}$.  By pumping enough
energy into the region one can create a black hole of area $A_{bh}
\leq A$, which means $S_{bh} \leq S < \bar S$.  On the other hand the
second law of thermodynamics requires $S_{bh} > \bar S$.  The only
solution is that there is no such state with entropy $\bar S$ and the
maximum entropy possible is precisely that of a black hole with
horizon area $= A$.
\bigskip

{\it In summary we see that the AdS/CFT conjecture is a precise and
explicit realization of the Holographic Principle and that the radial
dimension of AdS corresponds to a length scale in the theory on the
boundary of AdS}.
\bigskip

\noindent {\bf AdS/CFT correspondence rules}:
\bigskip

If the AdS/CFT conjecture is to serve a useful purpose then we need to
give a precise dictionary that relates processes in AdS to those in
the gauge theory.  Firstly we would expect that to a local gauge
invariant operator in the gauge theory there corresponds a field
propagating in AdS.  The boundary value of the field acts as a
source of the operator.  If we denote the field in AdS by $\phi(\vec
x,t,z)$ and the corresponding operator by ${\cal O}(\vec x,t)$, then to
first order the operator-source coupling is given by the interaction 
\be
S_I = \int d\vec x dt \ \phi(\vec x,t,z = \delta) {\cal O}(\vec
x,t)\delta^{\Delta - 4}.
\label{twentyone}
\ee
$\d{\lim_{\delta \rightarrow 0}} \ \phi(\vec x,t,\delta) = \phi_0(\vec x,t)$
is the boundary value of the AdS field $\phi(\vec x,t,z)$, and $\Delta$
is the scaling dim. of ${\cal O}(\vec x)$ and the correspondence of the two
theories is stated as:
\be
\langle \exp i \int d\vec x dt \delta^{\Delta-4} {\cal O}(\vec x,t) \phi_0(\vec
x,t)\rangle_{\rm SCFT} = {\cal Z}_{\rm string} \left(\phi(\vec x,t,\delta
\rightarrow 0) = \phi_0(\vec x,t)\right)
\label{twentytwo}
\ee
The l.h.s. is very precisely defined while the r.h.s.  
involves the full type IIB string theory partition function, which is
(at present) defined only in certain limits.  There is a whole class
of operators with special super symmetry properties for which we can
compute $\Delta$ in terms of the mass and spin of the field $\phi(\vec
x,t,z)$. 

\newpage

\noindent {\bf Tests of the AdS/CFT conjecture}:
\bigskip

To test eqn.(\ref{twentytwo}) we should be able to calculate both sides and
compare.  The strong form of the conjecture is that
eqn.(\ref{twentytwo}) is valid 
for finite $N$ and $g_s$.  The crucial ingredient that enables us to
test the conjecture are the identical symmetries of the gauge theory
and the string theory.  The ${\cal N} = 4$ gauge theory is invariant
under the super-conformal group $SU(2,2|4)$ whose bosonic sub-group is
$SO(4,2) \times SU(4)$.  $SO(4,2)$ is the conformal group in
4-dim.$\!\!$ and $SU(4)$ is the $R$-symmetry group corresponding to ${\cal
N} = 4$ supersymmetry.  $SU(2,2|4)$ is also a symmetry of the IIB
string theory with ${\rm AdS}_5 \times {\rm S}^5$ boundary
conditions.  We expect these symmetries to be valid for all values of
$g_s$ and $N$.  We can make further progress if we organize the gauge
theory in the 'tHooft $1/N$ expansion for fixed values of
$\lambda = g_s N$.  In this case, the basic relation $R/\ell_s = (4\pi
\lambda)^{1/4}$ implies that $\lambda \rightarrow 
\infty$ corresponds to $R/\ell_s \rightarrow \infty$.  In
this limit and as $N \rightarrow \infty$ the string theory can be
approximated by supergravity as an expansion in powers of $\ell_s/R$
and $\ell_p/R$.  $\ell_p$ is the Planck length 
defined by $\ell^8_p = \ell^8_s g^2_s$.
\bigskip

In supergravity one can perform a Kaluza-Klein reduction from ${\rm
AdS}_5 \times {\rm S}^5$ to ${\rm AdS}_5$.  For example a scalar field
$\Phi(x,y)$ can be expanded as $\Phi(x,y) = \d{\sum^\infty_{k=0}}
\phi^k(x) Y_k(y)$ where $x$ is a coordinate in AdS$_5$ and $y$ is a
coordinate in ${\rm S}^5$.  $Y_k(y)$ are the scalar sperical hermonics
on ${\rm S}^5$ and are given by the symmetric tensors $Y_k(y) \sim
y^{I_1} y^{I_2} \cdots y^{I_k}$. $y^I$ are the coordinates of a unit
vector on ${\rm S}^5$.  From the wave eqn. for $\Phi(x,y)$ we
can infer the mass of the field $\phi^k(x)$ in ${\rm AdS}_5$.  It
turns out to be $R^2 m^2_k = k(k+4)$ where $k(k+4)$ is the value of
the Casimir in the $(0,k,0)$ representation of $SO(6) \sim SU(4)$.  On
the Yang-Mills side the fields $\phi^k(x)$ correspond to traceless
symmetric tensors of $SO(6)$ formed out of the operator ${\cal O}^{I_1
\cdots I_k} = {\rm Tr}(\phi^{I_1} \cdots \phi^{I_k})$.  Those
operators also transform in the $(0,k,0)$ representation of $SO(6)$
and have dimension $k$.  Hence the formula $R^2m^2_k = k(k+4)$ implies
$k = \Delta = 2 + (4 + R^2 m^2_k)^{1/2}$, a relation between the
dimension $\Delta$ of an operator in the gauge theory and the mass of
the corresponding field in string theory.
\bigskip

In a similar way one can achieve a complete correspondence of
\underbar{all} supergravity short multiplets in AdS$_5$ and chiral
primary operators in the gauge theory.  A crucial point about the
correspondence is that the formulas that relate dimensions of the
gauge theory operators and masses of the supergravity fields are
independent of $\lambda$, and hence are valid at both strong and weak
couplings.
\bigskip

The AdS/CFT correspondence can also be tested by an exact ($\lambda$
independent) matching of anomalous terms: the non-abelian anomaly of
the $SU(4)$ $R$-symmetry currents is exactly reproduced by a
corresponding $SU(4)$ Chern-Simons term in the string theory.  It also
turns out that the 3-point functions of chiral primary operators in
the gauge theory exactly match the calculation from the string theory
side.
\bigskip

\noindent{\Large\bf 7 Uses of the AdS/CFT conjecture} \cite{one,three}: 
\bigskip

Given the physical basis, the maximal superconformal symmetry, and the
spectral correspondence we sketched in the previous section, we believe
that the conjecture is on a firm footing and can be used to derive
complementary results and insights on both sides of the correspondence.
\bigskip

\noindent {\bf Heavy quark potential at strong coupling}:
\bigskip

Let us first discuss an example of a strong coupling calculation in
the gauge theory using the correspondence.  We wish to calculate the
energy $E(L)$ of two heavy quarks, in the fundamental representation of
$SU(N)$, separated by a 
distance $L$ in the gauge theory.  The correct operator whose
expectation value evaluates this energy is a generalization of the
standard Wilson loop because the `heavy quarks' couple both to the
gauge fields and the scalar fields $\phi^I$.
\be
W(C) = {\rm Tr}\left[P \exp\left(\oint\left(i A_\mu {dx^\mu \over
d\tau} + y_I \Phi^I \sqrt{{\dot x}^2}\right)d\tau\right)\right]
\label{twentythree}
\ee
$x^\mu(\tau)$ defines the loop in space-time.  $y_I$ is a unit
vector on $S^5$ and ${\dot x}^2 = {\dot x}_\mu {\dot x}^\mu$.  
In the limit where $R/\ell_s \rightarrow \infty$, $\langle
W(C)\rangle$ can be represented as a first quantized string path
integral in AdS$_5 \times {\rm S}^5$, with boundary conditions at the
curve $C$.  It can be evaluated semi-classically for $\lambda \gg 1$,
by a surface that minimizes the path integral.  For a loop $C$
specified by the distance $L$ and time $T$, we have $\langle W \rangle
\approx e^{-TE(L)}$ and this gives 
\be
E(L) = -{4\pi^2 \over \Gamma\left({1\over4}\right)^4} {\sqrt{2\lambda}
\over L}
\label{twentyfour}
\ee
The fact that $E(L) \propto 1/L$ can be inferred from
conformal invariance.  It is the $\sqrt{\lambda}$ dependence that is
the prediction of AdS/CFT and it differs from the perturbative
estimate of $E(\lambda) \propto \lambda/L$.
\bigskip

\noindent {\bf Thermal gauge theories at strong coupling and black 
holes} \cite{ten}:
\bigskip

The ${\cal N} = 4$, super Yang-Mills theory defined on $S^3 \times
R^1$ can be considered at finite temperature if we work with euclidean
time and compactify it to be a circle of radius $\beta = 1/T$, where
$T$ is the temperature of the gauge theory.  We have to 
supply boundary conditions which are periodic for bosonic fields and
are anti-periodic for fermions.  These boundary conditions break the
${\cal N} = 4$ supersymmetry, and the conformal symmetry.  However the
AdS/CFT conjecture continues to hold and we will discuss the
relationship of the thermal gauge theory with the physics of black
holes in AdS.
\bigskip

As we have mentioned in the limit of large $N$ (i.e. $G_N \ll 1$) and
large $\lambda$ (i.e. $R \gg \ell_s$), the string theory is well
approximated by supergravity, and we can imagine considering the
Euclidean string theory partition function as a path integral over all
metrics which are asymptotic to AdS$_5$ space-time.  (For the moment we
ignore $S^5$).
\bigskip

The saddle points are given by the solutions to Einstein's equations
in 5-dim. with a -ve cosmological constant
\be
R_{ij} + {4 \over R^2} g_{ij} = 0
\label{twentyfive}
\ee
As was found by Hawking and Page, a long time ago, there are only two
spherically symmetric metrics which satisfy these equations with
AdS$_5$ boundary conditions: AdS$_5$ itself and a black hole solution.
The metric for both solutions can be written as
\bea
ds^2 &=& -V(r) dt^2 + V^{-1}(r) dr^2 + r^2 d\Omega^2_3 \nonumber \\
&& \\[2mm]
V(r) &=& 1 + {r^2 \over R^2} - {\mu \over r^2} \nonumber
\label{twentysix}
\eea
$\mu = 0$ corresponds to AdS$_5$ and $\mu > 0$ leads to a horizon
radius $r_+$ given by $V(r_+) = 0$.  The temperature of the black hole
is  
\be
{1 \over TR} = {2\pi R r_+ \over 2r^2_+ + R^2}
\label{twentyseven}
\ee
Let us denote the $\mu = 0$ solution by $X_1$ and the $\mu > 0$
solution by $X_2$.  These two spaces are topologically distinct
because in $X_2$ the boundary circle can be shrunk to zero while in
$X_1$ that is not possible.  In fact this property of $X_2$ 
defines a Euclidean black hole.  As a function of temperature
(\ref{twentyseven}) has infact 2 roots and we will choose the largest
root as it corresponds to a black hole with positive specific heat.
Note that for $r_+ \gg R$, $TR \sim \d{r_+ \over R}$. 
\bigskip

We can also calculate the euclidean Einstein-Hilbert action of $X_1$
and $X_2$.  Since both actions are infinite we can calculate both of
them using an appropriate cut off and then evaluating the difference
\be
I(X_2) - I(X_1) = {\pi^2 r^3_+(R^2 - r^2_+) \over 4G_5(2r^2_+ + R^2)}
\label{twentyeight}
\ee
The important point is that there occurs a change of dominance from
$X_1$ to $X_2$ at $r_+
= R$ and for $r_+ > R$, $I(X_2) < I(X_1)$.  The temperature at $r_+ =
R$ is $TR = \d{3 \over 2\pi} \sim o(1)$.  Hawking and Page interpreted
this as a first order phase transition from a phase consisting of
thermal gravitons to a large black hole.  For $r_+ \gg R$,
(\ref{twentyeight}) becomes
\be
I(X_2) - I(X_1) = -{\pi^5 \over 8} (TR)^3 {1 \over R^{-3} G_5} =
-{\pi^5 \over 8} N^2(RT)^3
\label{twentynine}
\ee
where we have used $R^5 G_5 = G_{10} = g^2_s \ell^8_s$ and $\d{R \over
\ell_s} = (g_s N)^{1/4}$.
\bigskip

Now let us discuss the above phenomenon in dual gauge theory at finite
temperature.  The 
${\cal N} = 4$ Super Yang-Mills theory at finite temperature involves
a continuation to periodic Euclidean time $\tau \sim \tau + 2\pi\beta$.
Hence the theory is defined on $S^3 \times S^1$.  A well known order
parameter at finite temperature is the Polyakov line defined as
\be
P(\vec x,\beta) = {\rm Tr}\left[P \exp \left(\int^\beta_0 d\tau
A_0(\vec x,\tau)\right)\right]
\label{thirty}
\ee
Now $SU(N)$ has a non-trivial centre given by $Z_N =
\left\{e^{\d{i2\pi k \over N}}, \ k = 1,\cdots,N\right\}$.  All local
gauge invariant operators are invariant under $Z_N$.  However the order
parameter (\ref{thirty}) transforms as 
\be
P'(\vec x,\beta) = g P(\vec x,\beta), \ g \in Z_N.
\label{thirtyone}
\ee
The phase in which $\langle P \rangle = 0$, $Z_N$ symmetry is intact.
It is called the `confinement phase' because the dominant excitations
in this phase correspond to color singlet single trace matrix products
of the fields of the gauge theory.  Denoting a typical $SU(N)$ matrix
valued field by $M$ these excitations correspond to ${\rm Tr}(M_{i_1}
M_{i_2} \cdots M_{i_n})$, where the length of the `string' $n < N^2$.
The free energy in this phase $F \sim N^0$.
\bigskip

In the phase $\langle P \rangle \neq 0$, $Z_N$ symmetry is broken.
This is called the `deconfinement phase', because the $N^2$ `color'
degrees of freedom are deconfined and the free energy $F \sim o(N^2)$.
In fact by using the fact that the underlying theory is conformally
invariant the free energy in this `deconfinement phase' is given by
\be
F \sim N^2 (RT)^3
\label{thirtytwo}
\ee 
This answer matches eqn.(\ref{twentynine}).  Since the gauge
theory is strongly coupled it is not possible to compute the numerical
coefficient in (\ref{thirtyone}), however on the gravity side it can
be computed!
\bigskip

The conclusion we can draw from the agreement of (\ref{twentynine})
and (\ref{thirtyone}) is that the `deconfinement' phase of the gauge
theory corresponds to the presence of a large black hole in AdS.
Besides qualitative predictions like (\ref{thirtyone}) it is difficult
to make precise quantitative statements about the gauge theory at
strong coupling $(\lambda \gg 1)$.  However on the AdS side the
calculation in gravity is semi-classical and hence there are precise
quantitative answers for 
\begin{enumerate}
\item[{(i)}] the temperature at which the first order
confinement-deconfinement transition occurs:
\[
T_c = {3 \over 2\pi R}
\]
\item[{(ii)}] the latent heat at $T = T_c$
\[
F(T_c) = N^2 {9\pi^2 \over 64}
\]
\item[{(iii)}] the free energy for $T > T_c$
\[
F(T) = N^2 {\pi^5 \over 8} (RT)^3
\]
\end{enumerate}
In the preceding 2 examples we saw how calculations in the strongly
coupled gauge theory can be done using the AdS correspondence by using
semi-classical gravity in the limit of $G_N \sim {1 \over N^2} \ll
1$ and ${R \over \ell_s} \sim \lambda^{1/4} \gg 1$.
\bigskip

We will now indicate how the correspondence can be used to make some 
non-trivial statements about the string theory (supergravity) in AdS.

\begin{enumerate}
\item[{(i)}] Just like in the case of the D1-D5 system, since the
${\cal N} = 4$ $SU(N)$ gauge theory is unitary, we can assert that
there cannot be information loss in any process in AdS, including
the process of the formation and evaporation of black holes.
However these processes need to be identified and worked out in this
gauge theory.

\item[{(ii)}] Another direction involved a study of the dynamics of small
Schwarzschild black holes in AdS.  Here when the horizon of the black
hole is of order the string scale, $r_h \sim \ell_s$, the supergravity
(space-time) description breaks down.  The string theory in this
region is not (yet) defined except by its correspondence with the
gauge theory.  The string size black hole of temp $T \sim \ell^{-1}_s$
is studied using a double scaling limit of the GWW
(Gross-Witten-Wadia) large $N$ phase transition in the gauge theory.
In the scaling region $T - T_c \sim N^{-2/3}$ a non-perturbative
description of a small string length size black hole was obtained
\cite{eleven}.
\end{enumerate}
\bigskip

\noindent {\bf Conformal Fluid dynamics and the AdS/CFT correspondence}:
\bigskip

We have seen that the thermodynamics of the strongly coupled gauge
theory in the limit of large $N$ and large $\lambda$ is calculable, in
the AdS/CFT correspondence, using the thermodynamic properties of a
large $(r_+ \gg R)$ black hole in AdS$_5$.  We now discuss how this
correspondence can be generalized to real time dynamics in this gauge
theory when both $N$ and $\lambda$ are large.  We will discuss a
remarkable connection between the (relativistic) Navier-Stokes
equations of fluid dynamics and the long wavelength oscillations of
the horizon of a black brane which is described by Einstein's
equations of general relativity in AdS$_5$ space-time.
\bigskip

On general physical grounds a local quantum field theory at very high
density can be approximated by fluid dynamics.  In a conformal field
theory in $3+1$ dim. we expect the density $\rho \propto T^4$, where
$T$ is the local temperature of the fluid.  Hence fluid dynamics is a
good approximation for length scales $L \gg 1/T$.  The
dynamical variables of relativistic fluid dynamics are the four
velocities: $u_\mu (x)$ $(u_\mu u^\mu = -1)$, and
the densities of local conserved currents.  The conserved currents are
expressed as local functions of the velocities, charge densities and
their derivatives.  The equations of motion are given by the
conservation laws.  An example is the conserved energy-momentum tensor
of a charge neutral conformal fluid:
\be
T^{\mu\nu} = (\epsilon + P) u^\mu u^\nu + P\eta^{\mu\nu} -
\eta\left(P^{\mu\alpha} P^{\nu\beta}(\partial_\alpha u_\beta +
\partial_\beta u_\alpha) - {1\over3} P^{\mu\nu} \partial_\alpha
u^\alpha\right) + \cdots
\label{thirtythree}
\ee 
where $\epsilon$ is the energy density, $P$ the pressure, $\eta$
is the shear viscocity and $P^{\mu\nu} = u^\mu u^\nu + \eta^{\mu\nu}$.
These are functions of the local temperature.  Since the fluid
dynamics is conformally invariant (inheriting this property from the
parent field theory) we have $\eta_{\mu\nu} T^{\mu\nu} = 0$ which
implies $\epsilon = 3P$.  Since the speed of sound in the fluid is
given by $v^2_s = \d{\partial P \over \partial \epsilon}$, $v_s = \d{1
\over \sqrt{3}}$ or re-instating units $v_s = \d{c \over \sqrt{3}}$,
where $c$ is the speed of light in vacuum.  The pressure and the
viscosity are then determined in terms of temperature from the
microscopic theory.  In this case conformal symmetry and the
dimensionality of space-time tells us that $P \sim T^4$ and $\eta \sim
T^3$.  However the numerical coefficients need a microscopic
calculation.
\bigskip

The Navier-Stokes equations are given by (\ref{thirtythree}) and
\be
\partial_\mu T^{\mu\nu} = 0
\label{thirtyfour}
\ee
The conformal field theory of interest to us is a gauge theory and a
gauge theory expressed in a fixed gauge or interms of manifestly gauge
invariant variables is not a local theory.  In spite of this
(\ref{thirtythree}) seems to be a reasonable assumption and the local
derivative expansion in (\ref{thirtythree}) can be justified using the
AdS/CFT correspondence.  
\bigskip

We now briefly indicate that the eqns.(\ref{thirtythree}),
(\ref{thirtyfour}) can be deduced systematically from black brane
dynamics.  Einstein's equation (\ref{twentyfive}) admits a boosted
black-brane solution 
\be
ds^2 = -2u_\mu dx^\mu dv - r^2 f(br)u_\mu u_\nu dx^\mu dx^\nu + r^2
P_{\mu\nu} dx^\mu dx^\nu
\label{thirtyfive}
\ee
where $v,r,x^\mu$ are in-going Eddington-Finkelstein coordinates and 
\bea
f(r) &=& 1 - {1 \over r^4} \nonumber \\ && \\
u^v &=& {1 \over \sqrt{1 - \beta^2_i}}, \ u^i = {\beta^i \over
\sqrt{1 - \beta^2_i}} \nonumber
\label{thirtysix}
\eea
where the temperature $T = 1/\pi b$ and the velocities
$\beta_i$ are all constants.  This 4-parameter solution can be
obtained from the solution with $\beta^i = 0$ and $b=1$ by  a boost
and a scale transformation.  The key idea is to make $b$ and $\beta^i$
slowly varying functions of the brane volume i.e. of the co-ordinates
$x^\mu$.  One can then develop a perturbative non-singular solution of
(\ref{twentyfive}) as an expansion in powers of $1/LT$.  Einstein's
equations are satisfied provided the velocities and pressure that
characterise (\ref{thirtyfive}) satisfy the Navier-Stokes
eqns. \cite{twelve}.  The pressure $P$ and viscosity $\eta$ can be
exactly calculated to be \cite{twelve,thirteen} 
\be
P = (\pi T)^4 \ {\rm and} \ \eta = 2(\pi T)^3
\label{thirtyseven}
\ee
Using the thermodynamic relation $dP = sdT$ we get the entropy density
to be $s = 4\pi^4 T^3$ and hence obtain the famous equation of
Policastro, Son and Starinets \cite{fourteen}, 
\be
{\eta \over s} = {1 \over 4\pi}
\label{thirtyeight}
\ee
which is a relation between viscosity of the fluid and the entropy
density.    
\bigskip

Systematic higher order corrections to (\ref{thirtythree}) can also be
worked out.  
\bigskip

{\it In summary we have a truly remarkable relationship between two famous
equations of physics viz. Einstein's equations of general relativity
and the relativistic Navier-Stokes equations}.  This relationship is
firmly established for a $3+1$ dim. conformal fluid dynamics which is
dual to gravity in AdS$_5$ space-time.  A similar connection holds for
$2+1$ dim. fluids and AdS$_4$ space-time.  A special (asymetric)
scaling limit of the relativistic Navier-Stokes equations, where we
send $v_s = \d{c \over \sqrt{3}} \rightarrow \infty$ leads to the
standard non-relativistic Navier-Stokes equations for an
incompressible fluid \cite{bhattacharya},\cite{fouxon}.
\bigskip

Finally it is hoped that the AdS/CFT correspondence lends new
insights to the problem of turbulence in fluids.  Towards this goal
the AdS/CFT correspondence has also been established for forced
fluids, where the `stirring' term is provided by an external metric and
dilaton field \cite{fifteen}.
\bigskip

In summary 
\bigskip

(i) the AdS/CFT correspondence allows us to discuss a
strongly coupled $(\lambda \gg 1)$ gauge theory at large density $(T
\gg 1)$ as a fluid dynamics problem whose equations are the
relativistic Navier-Stokes equations.  The various transport
coefficients and thermodynamic functions can be exactly calculated!
These results have encouraging implications for experiments involving
the collisons of heavy ions inspite of the fact that the relevant
gauge theory is QCD rather than ${\cal N} = 4$ Yang-Mills theory.  The
experiments at RHIC seem to support 
very rapid thermalization and a strongly coupled quark-gluon plasma
with very low viscosity coefficient.  There exists a window of
temperatures where the plasma behaves like a conformal fluid.  The AdS/CFT
correspondence also provides a calculational scheme for propagation of
heavy quarks and jet quenching in a strongly coupled plasma.
\bigskip

(ii) The Einstein equations enable a systematic determination of
higher derivative (in the velocities) terms of the Navier-Stokes
equations. 
\bigskip

(iii) The relationship between dissipative fluid dynamics and black
hole horizons, known as the membrane paradigm, has found a precise
formulation within the AdS/CFT correspondence.
\bigskip

(iv) The Navier-Stokes
eqns.(\ref{thirtythree}), (\ref{thirtyfour}) implies dissipation and
violates time reversal invariance.  The scale of this violation is set
by $\eta/\rho$ which has the dim. of length (in units where
the speed of light $c = 1$).  There is no paradox here with the fact
that the underlying theory is non-dissipative and time reversal
invariant, because we know that the Navier-Stokes equations are not a
valid description of the system for length scales $\ll \eta/\rho$,
where the micro-states should be taken into account.
\bigskip

\noindent {\bf QCD type theories and the AdS/CFT correspondence:}
\bigskip

QCD is not a conformally invariant theory.  As is well known at weak
coupling and at a length scale $L$, we have asymptotic freedom 
\[
Ng^2(L\wedge) = - {1 \over \beta_0 \ln L\wedge}, \ \beta_0 = {11 \over
24\pi^2} 
\]
$\wedge^{-1}$ is a fixed length that characterizes the theory.  In the
real world $\wedge \simeq 200$ mev, which corresponds to a length
scale $\simeq 10^{-13}$ cm.  For $L\wedge \gsim 1$, the theory is
strongly coupled and difficult to calculate.
\bigskip

Presently the best we can do is to study modifications of the ${\cal
N} = 4$ super-Yang-Mills theory, which we briefly mention.
\smallskip

\begin{enumerate}
\item[{a)}] Deforming the ${\cal N} = 4$ theory by `relevant'
operators can lead to new massive fixed points which are characterized
by a scale.  This can be established at strong coupling by seeking a
new supergravity solution that, in the interior of the
5-dim. space-time, differs from AdS$_5$.  The new solution breaks the
supersymmetry 
from ${\cal N} = 4$ to ${\cal N} = 1$ \cite{sixteen}.

\item[{b)}] One can wrap branes on cycles of space-time geometry that
break the supersymmetry.  The simplest example of this, is the wraping
of D4 branes on the thermal circle \cite{ten}.  The dual supergravity
background 
is quite simple as compared to a model in which D5-branes are wrapped
on a collapsed 2-cycle at a conifold singularity \cite{seventeen}. 
\end{enumerate}
\smallskip

Apart from reproducing qualitative features of non-perturbative
phenomena such as confinement, chiral symmetry breaking and the
low-energy spectrum of QCD, perhaps the most interesting qualitative
results that have been obtained in this approach are the QCD-like
behaviour in high energy and fixed angle (hard) scattering and the
qualitative properties of pomeron exchange.  These studies have
also yielded exciting connections of the Froissart bound in high
energy scattering with black hole physics \cite{eighteen}.
\bigskip

Finally we mention a novel application of the AdS/CFT correspondence
to gluon scattering amplitudes that employs a momentum space dual of
AdS$_5$ space-time \cite{nineteen}.
\bigskip

\noindent{\Large\bf 8 String Theory, Elementary Particles 
\smallskip

and Cosmology:} 
\bigskip

In the preceeding sections we have tried to present a case for string
theory as a framework to formulate a quantum theory of gravity.
String theory has passed many tests in this regard because it leads to
a perturbatively finite theory of gravity, and enables a calculation
of black hole entropy for a class of black holes.  This has been
possible primarily due to two important ingredients of string theory:
i) supersymmetry and ii) new degrees of freedom `$p$-branes' and in
particular $D$-branes.  At weak string coupling these are the solitons
of string theory.
\bigskip

The idea of supersymmetry requires that space-time has additional
fermionic co-ordinates besides the familiar bosonic co-ordinates, and
supersymmetric transformations mix these bosonic and fermionic
co-ordinates.  The existence of $D$-branes is also intimately
connected with the fact that the underlying theory is supersymmetric.
As we saw, in the discussion of black hole entropy, they are essential
for a unitary description of quantum gravity.
\bigskip

The microscopic understanding of black hole entropy and Hawking radiation
led to a precise formulation of holography, which may very well be one
of the guiding principles of string theory.  The Maldacena conjecture
gives a precise and calculable formulation of this idea.  Presently
there are various such dualities (AdS/CFT correspondence) known in
various space-time dimensions.  The most recent addition to this list
is a correspondence of a Chern-Simons gauge theory in 3-dim.$\!$ and $M_2$
branes which are objects of 11-dim.$\!$ $M$-theory \cite{nineteen-b}
\bigskip

The AdS/CFT correspondence seems to be useful for doing strong
coupling calculations not only in gauge theories but also for several
strongly coupled condensed matter systems see
e.g. \cite{nineteen-b,nineteen-c}.  Besides this it also led to a
duality between long wavelength motions on a black brane horizon and
dissipative fluid dynamics at high temperature and density.
\bigskip

Besides the unearthing of deep theoretical and
mathematical structures, we must ask whether the string theory
framework enables a description of elementary particle physics and
cosmology.  We briefly comment on this point.
\bigskip

\noindent {\bf Elementary Particle Physics:}\footnote{A modern
  reference is the book by M. Dine \cite{twenty}}
\bigskip

Our present understanding of elementary particle physics is based on
the Standard Model (SM) based on the gauge groups $SU(2) \times U(1)
\times SU(3)$.  $SU(2) \times U(1)$ describes the electroweak sector
and $SU(3)$ the strong interactions.  All particles of the SM, except
the Higgs, have been experimentally discovered and the SM gives a
precise description of elementary particle physics and the strong
interactions upto energies $\lsim 100$ GeV.  However the Higgs is not
yet discovered\footnote{The SM precision tests put a bound on the mass of
the Higgs: $M_H \gsim 114$ GeV.} and further there are 23
phenomenological parameters that are not calculable within the
standard model.  These parameters account for the masses of the quarks
and leptons, the sum of the Higgs, the masses of the neutrinos and the
small parameters that account for CP violation in the weak and strong 
interactions.
\bigskip

It is well known that while a non-zero expectation value of the Higgs
field accounts for electro-weak symmetry breaking and gives non-zero masses
to the quarks and leptons, its fluctuations lead to a quadratic
dependence on the ultra-violet cut off $\Lambda$, which make it
difficult to explain the heirarchy of the electro-weak scale ($\sim
100$ GeV) to the Planck scale ($\sim 10^{19}$ GeV): $M_{EW}/M_{Pl}
\simeq 10^{-17}$.  One of the proposed and robust mechanisms that
preserves this heirarchy, is supersymmetry.  The presence of
`fermionic loops' cancells the quadratic divergence!
\bigskip

The minimal supersymmetric standard model (MSSM) predicts a
unification of weak, electromagnetic and strong interactions at $\sim
10^{16}$ GeV, which is quite close to the Planck scale $\sim 10^{19}$
GeV.  This fact hints at a possible unification of the weak,
electromagnetic and strong interactions with gravity.  It is important
to note that this unification fails to happen in the absence of
supersymmetry.  Besides, unification, MSSM also provides a `dark
matter' candidate, viz. a neutralino, which is the lightest, stable,
neutral sypersymmetric fermionic particle.  It is `dark matter'
because it has only weak and gravitational interactions.  One of the
great expectations of the LHC (Large Hadron Collider), besides the
discovery of the Higgs particle, is the discovery of supersymmetry,
and a possible dark matter candidate!  Such a discovery would indeed
have profound consequences for both elementary particles and cosmology
\cite{twentyone}. Another proposed dark matter candidate is the axion.
\bigskip

Besides the MSSM there are other `effective field theory' proposals
involving higher dimensions, low scale supersymmetry, and warped
Randal-Sundrum type compactifications, that are being explored.  In
these scenarios the heirarchy problem disappears because the string
scale (or equivalently the 10-dim. Planck scale) is brought down to 1
TeV, so that quantum gravity effects can become experimentally
accessible at accelerators.  In the Randal-Sundrum scenario the
Standard Model lives on a 3-brane while only gravity extends in the
additional six dimensions.  The heirarchy of scales, or the smallness
of the weak scale as compared to the Planck scale, is explained in
terms of a gravitational red shift that occurs from a Planck brane
situated in the additional dimension.
\bigskip

We will not discuss the details of the various `Beyond the
Standard Model' proposals.  Fortunately the LHC, which goes into
operation a few 
months from now, will in the next few years give (hopefully!) a
verdict about nature's choice.  For the sake of argument let us
optimistically assume that supersymmetry is discovered at the LHC.  It
is likely that most of the physics below 1 TeV can be described in
terms of an effective 4-dim. lagrangian, with a partial list of
parameters fitted from experiment.  It would be difficult to find
direct evidence of string theory in a `low energy' effective
lagrangian unless we can calculate its parameters from an underlying
theory.  The issue is similar to asking whether we can infer the
details of the atomic and molecular composition of a fluid knowing the
fluid dynamics equations and the various coefficients like viscosity
that enter the equations.
\bigskip

In spite of the difficulty in identifying a signature of string theory
at low energies, 
the discovery of supersymmetry at the LHC would be an encouraging
sign about the string theory framework.  The difficult theoretical problem is
the exact emergence of the MSSM (or its variants) from string theory.
Brane constructions, compactification on manifolds with singularities
do come close to delivering the MSSM, within the framework of a theory
which also includes a consistent quantum theory of gravity.  This
point may have a consequence for one of the most puzzling facts of
nature, viz. the cosmological constant is a small but non-zero
number which is equivalent to a vacuum energy density of $10^{-8} \
{\rm erg/cm}^3$.
\bigskip

\noindent {\bf Cosmology:} \cite{twentytwo,twentythree,twentyfour}
\bigskip

The so called `standard model of cosmology' is not as well developed
as the standard model of particle physics.  It is an effective theory
with fewer parameters and its basic equations are Einstein's equations
of general relativity, where the stress tensor depends upon the matter
composition of the universe.  Even though there is no concrete
dynamical model of inflation (there are proposals) it is a key idea
that reconciles the Big Bang hypothesis with a large, homogeneous and
isotropic universe with fluctuations that eventually led to the
formation of matter and galaxies over a period of 14 billion years.
\bigskip

The large volume of experimental data from the cosmic micro-wave background
(CMB) radiation seems to be consistent with a `flat' homogeneous and
accelerating universe.  Another important fact is that only 4\% of the
energy density of the universe is the matter that we are familiar with
viz. the matter content of the standard model of particle physics.
The next abundant 22\% energy density source comes from `dark matter',
which is electrically neutral and hence optically dark.  The rest of
the 76\% is `vacuum energy' parametrized in the Einstein theory by the
cosmological constant.
\bigskip

There are several very basic questions we do not have answers to and
which are subjects of active research.  We briefly comment on these.

\begin{enumerate}
\item[{i)}] Inflation is usually modeled by a single scalar field
called the inflaton.  String theory which seems to be a compelling
ultra-violet completion of general relativity does not lead to a
single scalar field but to a whole host of scalar fields (moduli
corresponding to shapes and sizes of the internal compact newfold) which
naturally occur in string compactifications.  The moduli stabilization
problem is the same as giving a mass to the corresponding scalar
fields.  One possibility is to turn on discrete fluxes in the 6
compact directions of string theory.  This leads to, in models that
have been studied, to a lifting of all but the `size' or volume
modulus, and the potential as a function of this modulus can be
calculated.  The minimum turns out to 
be an AdS space-time.  In order to break the supersymmetry and raise
the ground state energy by a small amount, a probe anti-D3 brane is
introduced.  In this way a meta-stable vacuum with positive vacuum
energy is achieved. This leads to a de
Sitter space-time with a non-zero and positive cosmological constant.
The above construction is called the Kachru, Kallosh, Linde and
Trivedi (KKLT) scenario \cite{twentyfive}.  This and various other
$D$-brane inflation 
scenarios are a subject of active investigations.  It is fair to say
that presently, the `slow roll' or prolonged inflation seems to be difficult
to achieve.  Perhaps progress in locating the standard model vacuum in
string theory may contribute to a solution of this very important and
difficult problem.
\item[{ii)}] The standard model of cosmology has the inevitable space
like `cosmological' singularity, as one scales the size of the
universe to zero.  Here general relativity breaks down, and string
theory should certainly be relevant for a resolution of this
singularity.  There are various attempts in this direction within
effective field theory, perturbative string theory and matrix models.
What would be desirable is to find a model cosmology in AdS$_5$ and
study its hologram in the gauge theory on the boundary of AdS$_5$.  A
resolution of the space like singularity of a black hole is bound to 
shed light on this fundamental question.
\item[{iii)}] As we mentioned in (i), the turning on fluxes in the
compact manifold leads to a large number $(\sim 10^{500})$ of
acceptable ground states of string theory. The question arises whether
there is a drastic reduction of these consistent ground states
in the full non-perturbative theory.  We do not know the answer to
this question, because we do not know non-perturbative string theory
well enough except in the case of AdS space-times, where it is dual to
a $SU(N)$ gauge theory.  However one cannot precude the possibility
that our universe, which presently has a vacuum energy density
$\rho_{rac} \simeq 10^{-8} \ {\rm erg/cm}^3$, is not special!  As
Weinberg \cite{twentyeight-b} noted, if $\rho_{rac} > 10^{-8} \ {\rm
  erg/cm}^3$, we would 
have an universe in which galaxies could not have formed.  The key
question is whether such an uninteresting universe is also a consistent
solution of `string theory'?
\end{enumerate}
\bigskip

In summary it is fair to say that we presently do not know string
theory well enough both conceptually and technically to provide
answers to questions and issues we have discussed above.  It seems
certain that we will need to explore models in which both cosmology
and particle physics are tied up in a dynamical way.
\bigskip

\noindent {\Large\bf 9 Epilogue}
\bigskip

In this review we have made the case for string theory as a frame work
for a finite theory of quantum gravity that goes beyond quantum field
theory.  Along the way to realize this,  
a host of new and intricate structures have been discovered.
Prominent among these are supersymmetry and the holographic
principle.  The raison d'etre of string theory is elementary particle
physics and cosmology, and the quest to answer basic questions
related to the fundamental structure of matter and the laws of the
cosmos.  However like all fertile ideas in science string theory and
its methods make a connection with other areas of physics and
mathematics.  Its influence on geometry and topology is well known
\cite{twentysix}\footnote{Also see M.F. Atiyah's article on `Einstein
and Geometry in \cite{thirtythree}.}.  Its ability to solve
outstanding strong coupling 
problems in condensed matter physics is being realized in the AdS/CFT
correspondence.  The connection with fluid dynamics is also
tantalizing and may perhaps shed light on the problem of turbulence. 
\bigskip

A more popular exposition of the topics we have discussed here can be
found in \cite{thirtythree}.  Especially relevant are the articles by
D. Gross, M.F. Atiyah, A. Sen, A. Dabholkar and S. Sarkar.
\bigskip

\noindent {\large\bf Acknowledgement:}
\bigskip

I would like to thank Avinash Dhar, Gautam Mandal, Shiraz Minwalla and
Sandip Trivedi for many insightful and enjoyable discussions.

\end{document}